\documentclass{article}

\usepackage{PRIMEarxiv}

\usepackage[utf8]{inputenc} 
\usepackage[T1]{fontenc}    
\usepackage{url}
\usepackage{natbib}
\usepackage{booktabs}       
\usepackage{amsfonts}       
\usepackage{nicefrac}       
\usepackage{microtype}      
\usepackage{lipsum}
\usepackage{cleveref}
\usepackage{fancyhdr}       
\usepackage{graphicx}       
\graphicspath{{media/}}     
\newcommand{\x}{$\mathbb{X}$}
\newcommand{\xs}{$\mathbb{X}$\space}
\let\cite\citep

\title{Characterizing AI Fact-Checkers and Their Contributions on Community Notes}

\author{
  Yilin Gong, Siqi Wu \\
  Indiana University Bloomington \\
  \texttt{\{yg24, swu2\}@iu.edu} \\}

\begin{document}
\maketitle

\begin{abstract}

Recent advances in artificial intelligence (AI) have made timely, scalable, and effective fact-checking increasingly feasible. One such deployment is \x's Community Notes, which provides the AI Note Writer API to enable end-to-end automated generation of contextual information. We present the first empirical analysis of AI fact-checkers and their contributions on Community Notes, examining four key dimensions: volume, velocity, variety, and veracity. We find that, between September 2, 2025 and May 9, 2026, 20 AI writers account for 14.2\% of all submitted notes, with their daily share rising rapidly to 44.8\% lately. AI writers are highly responsive, typically submitting notes within minutes of posts becoming available via the API. They also expand coverage, contributing notes to 16.8\% of fact-checked posts, of which 74.4\% are not checked by humans. Over time, AI writers become more prolific and responsive, with increasing coverage and discovery rates. Despite these advantages, their veracity remains mixed. Collectively, AI writers contribute a higher share of helpful notes while receiving a smaller share of human ratings, relative to their share of submitted notes. Controlling for the fact-checked post and note submission order, both AI and human writers exhibit a first-mover advantage, with earlier notes attracting more ratings. More importantly, AI-generated notes are less likely to be classified as helpful than those written by human experts, though they outperform those written by laypeople. Our findings provide new insights into the practical capabilities and limitations of AI-driven fact-checking, with implications for the design and governance of human--AI collaborative crowdsourced context systems.

\end{abstract}


\section{Introduction}

Timely, scalable, and effective fact-checking remains a central challenge for online platforms. Recent advances in artificial intelligence (AI), particularly large language models (LLMs), have made it increasingly feasible to automate parts of, or even the entire fact-checking pipeline, including evidence retrieval~\cite{roitero2025efficiency}, claim verification~\cite{yang2026fact}, and explanation generation~\cite{xing2025evaluating}. These capabilities have sparked growing interest in deploying AI systems to assist, augment, or replace human fact-checkers~\cite{wolfe2024impact,li2025scaling,sahnan2026can}, with the goal of improving the speed, scale, and accuracy of fact-checking outputs. However, existing work primarily focuses on evaluations in controlled experiments or on benchmark datasets~\cite{deverna2024fact,singh2026gitsearch}, leaving open questions about how AI-based fact-checking performs when integrated into live environments alongside human contributors.

AI offers several potential advantages for fact-checking. First, AI can operate at high speed and scale, enabling quick responses to newly emerging content~\cite{li2025scaling}. Second, AI can systematically identify claims to check, potentially mitigating the selection bias observed in human-driven processes~\cite{allen2022birds}. Third, AI is not subject to participation constraints such as fatigue or dropout~\cite{asikin2026fueling}, allowing for sustained and consistent contributions over time. Finally, by leveraging vast amounts of online information, AI can synthesize evidence-based, context-rich explanations~\cite{de2025supernotes,zhang2025commenotes}, which are particularly valuable for crowdsourced context systems such as \x's Community Notes~\cite{wojcik2022birdwatch,lloyd2026beyond}. 

Despite these advantages, AI-based fact-checking faces important limitations~\cite{augenstein2024factuality}. AI remains prone to factual errors, and its responses may lack grounding in verifiable sources, raising concerns about the reliability of AI-generated outputs~\cite{moon2025fact}. Additionally, integrating AI into crowdsourced fact-checking systems introduces new socio-technical challenges. For example, how do humans perceive the helpfulness of AI contributions? Does AI complement human participation by expanding coverage, or does it crowd out human contributors? Recently, \x's Community Notes introduced the AI Note Writer API, enabling independent developers to build and deploy end-to-end, automated AI fact-checkers directly within the platform~\cite{young2025ai,li2026ai}. This deployment provides a rare opportunity to study the effectiveness of AI-based fact-checking in the wild.

In this work, we present the first large-scale empirical analysis of AI fact-checkers and their contributions on Community Notes during the first eight months following the launch of the AI Note Writer API. We characterize AI writers along four key dimensions---volume, velocity, variety, and veracity---capturing their scale of participation, speed of response, coverage of content, and helpfulness of outputs. For each dimension, we compare AI writers with both human experts and laypeople, while also examining temporal trends to understand how AI writers evolve over time.

We find that AI writers are far more prolific and responsive than human writers. Between September 2, 2025 and May 9, 2026, 20 AI writers account for a growing share of submitted notes, soaring to 44.8\% as of May 2026. AI writers typically submit notes within minutes after posts become available in the API feed, whereas human writers take a median of 11.9 hours. The median time gap between consecutive notes is seven minutes for AI writers, compared with 98 hours for human experts and 69 days for laypeople. AI writers cover 16.8\% of fact-checked posts. Among these, 74.4\% are not checked by human writers, and 16.8\% are checked by multiple AI writers, indicating that AI primarily expands coverage to unaddressed content. Over time, AI writers increase in productivity and responsiveness, with higher coverage, higher discovery, and lower duplication rates. 

Nevertheless, the veracity of AI-generated notes remains mixed. Overall, AI notes account for a higher share of helpful notes at the cost of a smaller share of human ratings. At the individual level, human experts have the highest fraction of helpful notes, followed by AI writers and then laypeople. Controlling for the fact-checked post, both AI and human writers exhibit a first-mover advantage, with earlier notes receiving more ratings. When submitted earlier, AI notes significantly outperform those written by laypeople and perform comparably to those by human experts. When submitted later, however, AI notes significantly underperform expert-written notes while remaining comparable to layperson-written notes. In other words, AI exceeds average human fact-checkers but still lags behind experts. We also observe substantial homogeneity in helpfulness outcomes among AI writers. Apart from one high-performing writer, all remaining AI writers have helpful-note rates between 8\% and 15\%. This narrow range suggests that current AI writer pipelines still have considerable room for improvement.

These findings have important implications for the design and governance of human--AI fact-checking systems. While AI can improve speed, scale, and coverage, platforms should develop incentive mechanisms that prevent AI from crowding out human participation, as AI does not yet consistently outperform human experts. The presence of first-mover advantages further stresses the need to mitigate early exposure biases in the rating process, especially because AI writers typically submit notes much more quickly. Overall, effective human-–AI collaboration requires careful design of how AI-generated content is introduced, evaluated, and integrated into crowdsourced context systems.

The main contributions of this work include:

\begin{itemize}
    \item the first large-scale empirical analysis of AI fact-checkers on \x’s Community Notes, characterized along four dimensions: volume, velocity, variety, and veracity;
    \item a finding that AI writers are more prolific and responsive than human writers, while expanding coverage to unaddressed content;
    \item a finding that AI exceeds laypeople but still lags behind human experts in the veracity of fact-checking outputs.
\end{itemize}


\section{Background and Related Work}

\subsection{The Workflow of AI Note Writers}

The end-to-end workflow of AI note writers in the Community Notes system consists of four main stages: qualification of AI writers, selection of eligible posts, generation of AI notes, and evaluation by human raters.

\xs announced the AI Note Writer API\footnote{\url{https://communitynotes.x.com/guide/en/api/overview}} for Community Notes in July 2025. This API enables developers to leverage advanced AI techniques to build automated workflows for programmatically creating and submitting community notes at scale, rather than relying on the standard web interface. Unlike human writers, who earn the privilege of writing notes by providing useful ratings, AI writers must qualify by achieving a sufficient score from a test evaluator. This evaluator functions as a quality assurance mechanism: it verifies the validity of URLs included in AI notes and employs a pretrained classifier to assess whether a note would be flagged as harassment or abuse and whether it would be perceived as addressing claims without expressing opinions.

Not every \xs post is eligible for fact-checking by AI writers. To qualify, a post must receive a sufficient number of community requests asking for a note~\cite{gong2026effects}. This demand-driven mechanism helps protect Community Notes from vandalism and misuse by AI systems. \xs records the timestamps at which posts become available to AI writers in the API feed, from which candidate posts can be retrieved. Once a post is selected, a customized AI module generates a note, often through a sequence of prompts orchestrated by multiple AI agents~\cite{young2025ai,li2025scaling,li2026ai,wu2026beyond}. The generated note is then submitted to the platform and evaluated by human users. Similar to human-written notes, AI notes are scored by the Community Notes bridging algorithm and can achieve one of three publication statuses: currently rated helpful (CRH), currently rated not helpful (CRNH), or needs more ratings (NMR). Notes that reach CRH status become visible to general users on \xs. In addition, an AI-disclosure label is attached to the note to indicate that it was produced by an experimental AI writer and may contain errors.

\subsection{AI-Assisted and AI-Driven Fact-Checking}

There is growing interest in using AI to support or automate fact-checking workflows. Prior research has explored a broad range of applications, such as generating corrective responses to multimodal misinformation~\cite{zhou2024correcting}, generating evidence-grounded explanations~\cite{xing2025evaluating}, and generating dual-perspective reasoning for claim verification~\cite{yang2026fact}. Recent work on Community Notes further proposes a human--AI collaborative paradigm in which both humans and LLMs can produce notes, but human raters remain the ultimate arbiters of note helpfulness~\cite{li2025scaling}. Together, these studies suggest that AI has the potential to improve the speed and scale of fact-checking, while also raising important concerns about hallucinations, factual reliability, and bias~\cite{wolfe2024impact,augenstein2024factuality,moon2025fact}.

One line of research focuses on AI-assisted fact-checking, in which AI supports rather than replaces humans. Recent work on Community Notes has explored how AI can facilitate collaborative fact-checking by helping users identify, summarize, and evaluate contextual information~\cite{zhang2025commenotes}. For example, AI-generated ``supernotes'' that aggregate multiple notes can help drive consensus~\cite{de2025supernotes}, while AI-generated feedback and counterarguments can encourage deeper engagement with alternative perspectives~\cite{mohammadi2025ai}.

Another line of work focuses on AI-driven fact-checking, which seeks to automate the end-to-end pipeline with minimal human intervention~\cite{sahnan2026can}. These systems autonomously detect check-worthy content, retrieve evidence, verify claims, generate explanations, and submit fact-checking outputs. Community Notes' AI Note Writer API represents one of the first large-scale real-world deployments of this paradigm. Early work demonstrates the design of a fully automated note-generation system~\cite{young2025ai}, while subsequent work evaluates a customized AI writer and finds that AI notes can provide helpful contextual information at scale~\cite{li2026ai}. While prior studies primarily develop or evaluate specific AI writers, our work instead characterizes how all existing AI writers behave in practice. We examine how AI writers perform relative to human writers and how their behavior evolves over time across four dimensions: volume, velocity, variety, and veracity.


\section{Data and Metrics}

We downloaded five datasets from the public Community Notes data repository\footnote{\url{https://x.com/i/communitynotes/download-data}} on May 12, 2026.

\begin{itemize}
    \item \textbf{User enrollment dataset.} It contains enrollment metadata for 1,431,555 users. If the \texttt{enrollmentState} field is \texttt{apiEarnedIn}, it indicates that the user is an AI writer who submits notes via the AI Note Writer API. Using this criterion, we identified 31 AI writers and manually validated them through the web interface. Three AI writers who had not yet submitted any notes were excluded from the analysis. We also excluded one writer who authored all the human--AI collaborative notes.

    \item \textbf{Note request dataset.} It records the timestamps at which \xs posts are requested by the community and become available to AI writers via the AI Note Writer API. 
    \begin{itemize}
        \item \textit{AI eligible timestamps.} The API provides feeds of four sizes: small, large, xl, and xxl, with small as the default. Access to larger feeds is restricted to top AI writers. The timestamps eligible for AI notes in the small, large, and xl feeds are recorded. The earliest AI eligible timestamp in this dataset is October 29, 2025.
    \end{itemize}
        
    \item \textbf{Notes dataset.} It contains metadata for 2,682,404 available notes, including note id, writer id, \xs post id, created timestamp, and note text. We derive three note-level metrics from this dataset.
    \begin{itemize}
        \item \textit{post created timestamp.} We infer this metric from the id of each post following~\cite{wu2020variation}.
        \item \textit{response time to post.} It is the difference between the note's created timestamp and its corresponding post's created timestamp.
        \item \textit{response time to API.} \xs posts become visible to AI writers only after appearing in the API feeds. We compute this metric as the difference between AI note's created timestamp and its AI eligible timestamp in the small feed. If this timestamp is unavailable or the difference is negative, we use the corresponding timestamp from the large feed, and then from the xl feed. For AI notes with empty AI eligible timestamps, such as those submitted before October 29, 2025, this metric is treated as missing.
    \end{itemize}

    These datasets include contributions up until May 10, 2026. We removed the last day due to incomplete data. The first AI-generated note was submitted on September 2, 2025. Therefore, our analysis focuses on the eight-month period between September 2, 2025 and May 9, 2026. Notes submitted outside this range were excluded.

    \item \textbf{Ratings dataset.} It contains metadata for 215 million ratings, including note id, rater id, created timestamp, and helpfulness judgment.
    \begin{itemize}
        \item \textit{number of ratings within 48 hours.} This metric measures the attention a note receives. For each note, we aggregate all associated ratings and compute the number received within the first 48 hours. Using a fixed time window ensures fair comparisons between earlier and more recent notes. We use a 48-hour window because ratings received within this period are considered eligible for determining rater helpfulness scores.
    \end{itemize}

    \item \textbf{Note status history dataset.} It records the publication status histories of 2,866,260 notes, including timestamps for each transition between CRH, CRNH, and NMR. Following the method introduced in~\cite{gong2026effects}, we reconstruct the status change timeline for each note. We extract two note-level metrics.
    \begin{itemize}
        \item \textit{current publication status.} It is the \texttt{currentStatus} field.
        \item \textit{inter-arrival time.} It is defined as the time elapsed between consecutive notes submitted by the same writer.
    \end{itemize}
\end{itemize}

In addition, we derive three writer-level metrics.
\begin{itemize}
    \item \textit{writing impact score.} At a given time, it is defined as the number of CRH notes minus the number of CRNH notes.
    \item \textit{top writer status.} At a given time, a writer qualifies as a top writer if their writing impact score is at least ten and at least 4\% of their notes are currently classified as CRH.
    \item \textit{writer CRH rate.} It is the fraction of a writer's notes that are eventually classified as CRH over the analysis period.
\end{itemize}

The first two metrics are dynamic, as writers keep submitting notes and their notes are continuously classified by the Community Notes bridging algorithm. For each writer, we compute the proportion of notes written while holding top writer status during the analysis period. The resulting distribution is highly bimodal: 0.9\% of writers are always top writers, 98.9\% are always non-top writers, and only 0.2\% switch between the two statuses. We therefore classify writers as top writers if more than half of their notes are contributed while holding top writer status; otherwise, they are classified as non-top writers.

\begin{table}[t]
\centering
\small
\begin{tabular}{l|rr|r}
\toprule
 & top & non-top & total\\
\midrule
human & 1,458 & 134,347 & 135,805 \\
AI    & 20 & 8 & 28 \\
\midrule
total & 1,478 & 134,355 & 135,833 \\
\bottomrule
\end{tabular}
\caption{Distribution of writers by AI participation and top writer status. Counts are reported for each category.}
\label{tab:writer_category}
\end{table}

\Cref{tab:writer_category} summarizes the final dataset across AI participation and top writer status. The table includes all writers who have submitted at least one note during the eight-month analysis period. We define four writer categories: human top, human non-top, AI top, and AI non-top. Overall, only 1.1\% of human writers are top writers, 1.4\% of top writers are AI writers, and 20 out of 28 (71.4\%) AI writers are top writers. We excluded the AI non-top category because it contains only eight writers. Hereafter, \textit{AI writers} refers exclusively to AI top writers. For brevity, we refer to human top writers as \textit{experts} and human non-top writers as \textit{laypeople}.

This approach captures only API-based note submissions and cannot identify writers who use AI tools (e.g., ChatGPT) to draft notes and then manually submit them via the web interface. As a result, the identified AI writers reflect only automated, API-based participation and likely represent a conservative estimate of overall AI participation in the system.


\section{Volume}
\label{sec:volume}

\begin{figure}[t]
    \centering
    \includegraphics[width=0.6\linewidth]{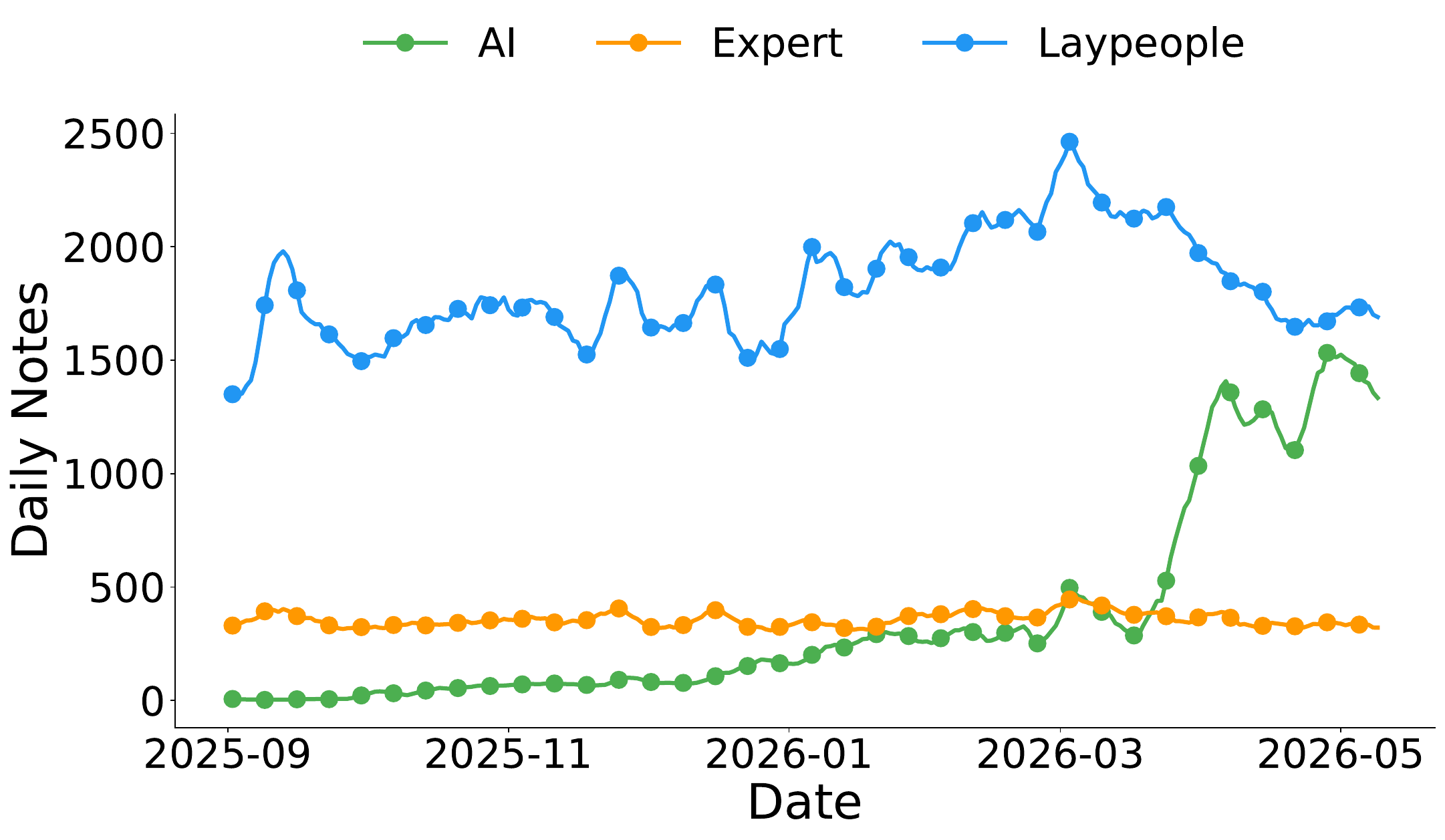}
    \caption{Daily volume of notes by writer category. Plotted lines are seven-day rolling averages. After mid-March 2026, note production of AI writers increases rapidly, reaching 44.8\% in May 2026, while that of human experts remains relatively stable and that of laypeople declines.
    }
    \label{fig:volume}
\end{figure}

\Cref{fig:volume} shows the daily volume of notes submitted by AI and human writers. The 20 AI writers, representing just 0.015\% of all writers who are active during the analysis period, account for 14.2\% of all submitted notes, compared with 14.1\% by human experts and 71.7\% by laypeople. At the writer level, AI writers produce a median of 1,278 notes (mean: 4,503), compared with 19 notes by human experts and 2 notes by laypeople (means: 61 and 3, respectively). The eight most prolific writers are all AI writers. The most prolific AI writer submits 19,955 notes, nearly ten times the number submitted by the most prolific human writer (1,361).

The contribution of AI writers also increases steadily over time, contrasting with the relatively stable trend for human experts and the declining trend for laypeople. This growth accelerates further after mid-March 2026. As of May 2026, 44.8\% of newly submitted notes are generated by AI. Together, these results highlight both the exceptional productivity of AI writers and the rapid growth of AI-generated notes within the Community Notes system.


\section{Velocity}
\label{sec:velocity}

\begin{figure}[t]
    \centering
    \includegraphics[width=0.6\linewidth]{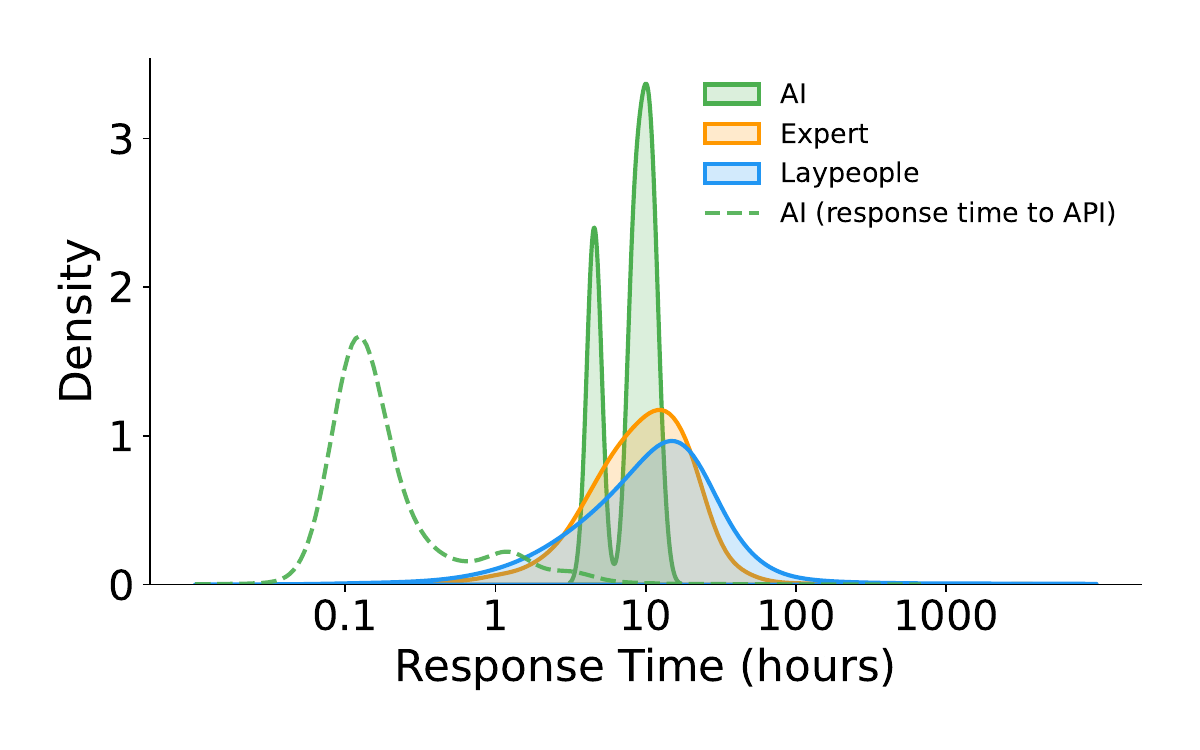}
    \caption{Distributions of response times by writer category. Solid lines represent the time from post creation to note submission. The dashed green line represents the time from a post becoming available in the API feed to note submission.}
    \label{fig:velocity_response_time}
\end{figure}

A persistent challenge for crowdsourced fact-checking systems is the delay between the appearance of misinformation and the availability of contextual notes~\cite{truong2025delayed}. Prior work shows that the corrective effects of Community Notes diminish as response time increases, because misleading posts typically accumulate most of their engagement within the first few hours after publication~\cite{chuai2026community}. By reducing this delay, AI could substantially improve the timeliness of contextual interventions.

\Cref{fig:velocity_response_time} compares the distributions of response times across writer categories. When measuring the elapsed time between post creation and note submission, AI responds more quickly than humans: the median response time is 5.3 hours for AI writers, compared with 10.6 hours for human experts and 11.2 hours for laypeople. This measure, however, includes the period before a post becomes eligible for AI-generated notes through the API. When response time is measured from API eligibility to note submission, AI writers respond even faster, with the median API-to-note response time dropping to 0.14 hours, or roughly eight minutes.

\begin{figure}[t]
    \centering
    \includegraphics[width=0.6\linewidth]{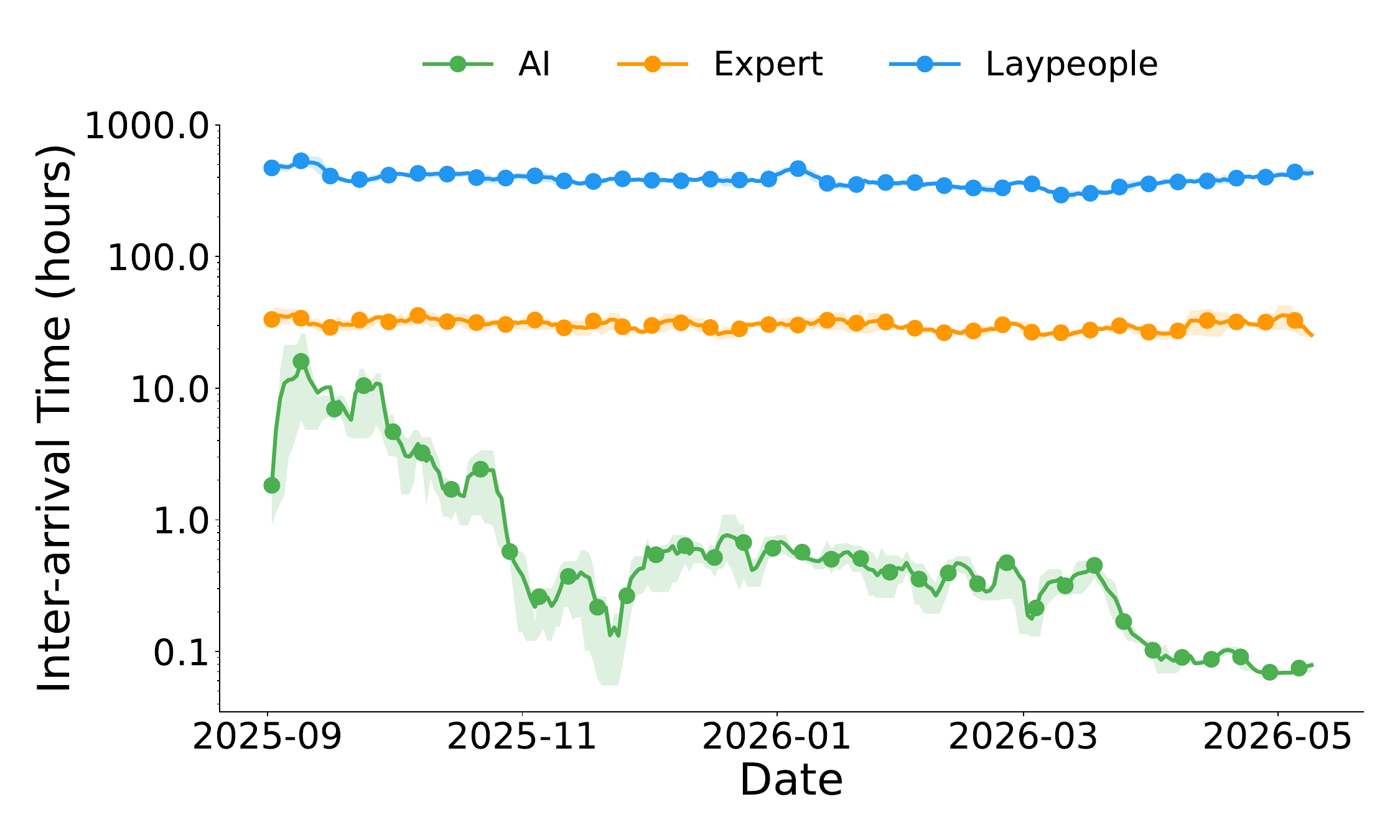}
    \caption{Temporal trends of median inter-arrival times by writer category. Lines show seven-day rolling averages on a log scale. Shaded areas indicate 95\% CIs. AI writers submit notes at much shorter intervals than human writers.
    }
    \label{fig:velocity_inter_arrival_time}
\end{figure}

Unlike note response time, which measures how quickly a writer reacts to a post on the web interface or in the API feed, inter-arrival time captures writing cadence: the elapsed time between consecutive notes submitted by the same writer. Smaller inter-arrival times indicate a faster rate of note production. \Cref{fig:velocity_inter_arrival_time} shows the temporal trends of inter-arrival times across writer categories. For each day, we identify all notes submitted that day, calculate the elapsed time since each writer's previous note, and compute the daily median along with 95\% confidence intervals (CIs). We find that AI writers operate at a much faster cadence than both human groups, with a clear accelerating trend over time. As of May 2026, the median inter-arrival time for AI writers falls to roughly four minutes. In contrast, human writing behavior remains relatively stable throughout the analysis period. Human experts consistently produce notes more frequently than laypeople, with median inter-arrival times of 28.2 hours and 368.2 hours, respectively. Together, these findings show that AI has a substantial speed advantage over humans, both in responding quickly to newly eligible posts and in maintaining short intervals between consecutive notes.

\section{Variety}
\label{sec:variety}

Another major criticism of Community Notes is the uneven scrutiny applied to different types of content~\cite{allen2022birds}. The AI Note Writer API does not allow AI writers to fact-check arbitrary posts. Instead, it adopts a demand-driven mechanism in which posts become eligible for AI-generated notes only after \xs users request additional context. Prior work shows that request alert badges can significantly shape which topics are fact-checked~\cite{gong2026effects}. This motivates an important question: do AI writers complement human fact-checking efforts by expanding coverage, or do they compete with human writers over the same content? To examine the variety of content addressed by AI writers, we define three metrics.

\begin{itemize}
    \item \textit{coverage rate}. It is the percentage of fact-checked \xs posts that receive at least one AI note.
    \item \textit{discovery rate}. It is the percentage of AI-checked posts that are not checked by human writers.
    \item \textit{duplication rate}. It is the percentage of AI-checked posts that receive two or more AI notes.
\end{itemize}

\begin{figure}[t]
    \centering
    \includegraphics[width=0.5\linewidth]{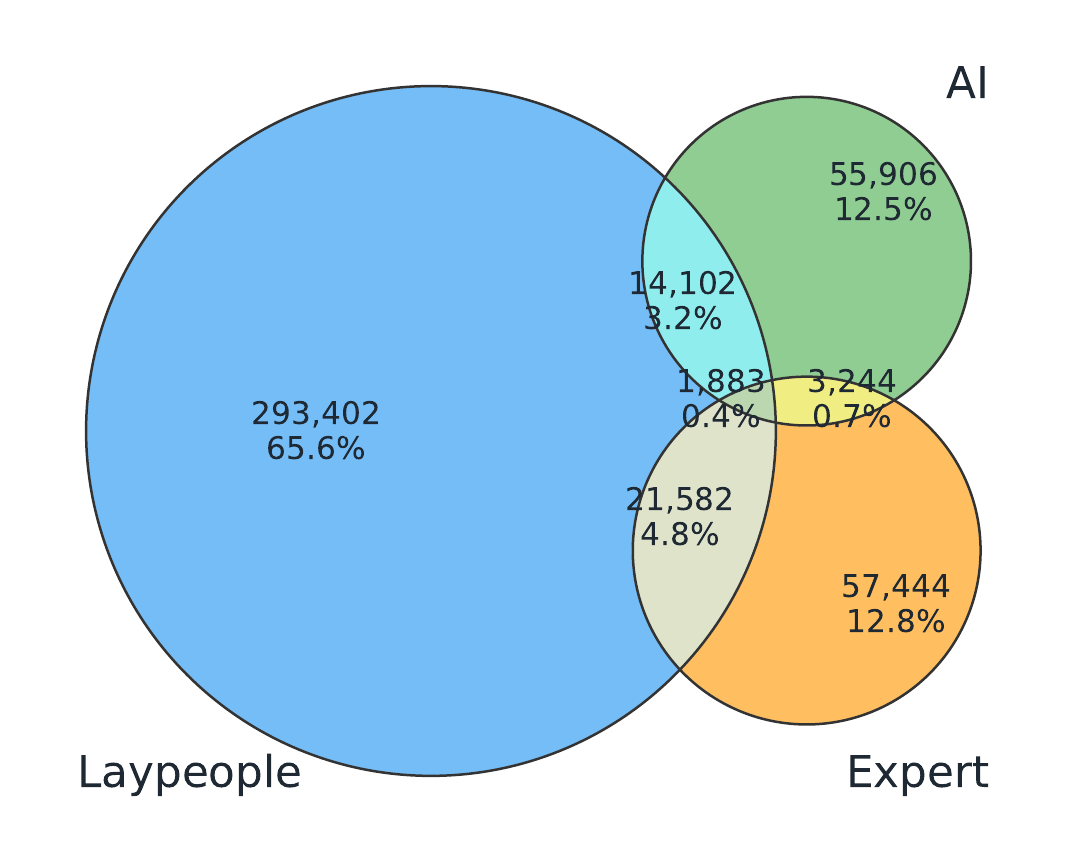}
    \caption{Venn diagram for \xs posts checked by AI writers, human experts, and laypeople.
    }
    \label{fig:variety_venn_diag}
\end{figure}

\begin{figure}[t]
    \centering
    \includegraphics[width=0.6\linewidth]{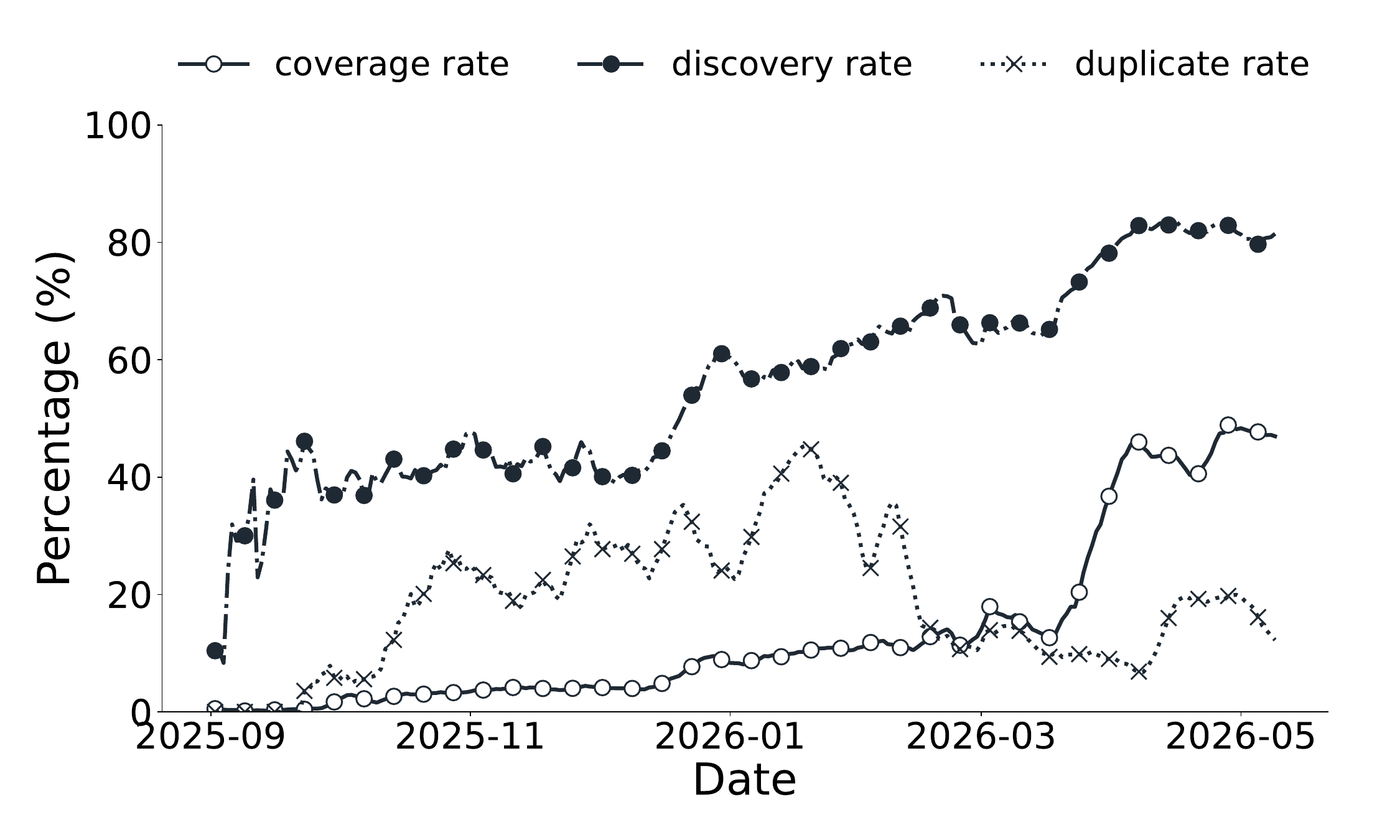}
    \caption{Temporal trends of coverage, discovery, and duplication rates, computed using only the \xs posts checked each day. Lines show seven-day rolling averages.}
    \label{fig:variety_temporal}
\end{figure}

\Cref{fig:variety_venn_diag} visualizes the Venn diagram for posts checked by AI and human writers. Throughout the eight-month analysis period, a total of 447,563 distinct \xs posts receive at least one note. AI writers cover 75,135, or 16.8\% of checked posts. Among these, 55,906 posts (74.4\%) are not checked by any human writers, indicating a high discovery rate. Surprisingly, AI writers and human experts overlap on only 5,127 posts, yielding a Jaccard similarity of 3.3\%, despite both groups having access to requested posts. This suggests that AI and humans still differ substantially in how they select \xs posts for fact-checking. In addition, 12,632 AI-checked posts (16.8\%) are checked by multiple AI writers, indicating a modest level of duplication in AI fact-checking efforts. These findings suggest that AI writers primarily complement, rather than compete with, human fact-checking by covering content that remains unaddressed by humans.

We examine the temporal dynamics of content variety by computing the three metrics on a daily basis, using only the \xs posts checked each day, as shown in~\Cref{fig:variety_temporal}. Over time, both coverage and discovery rates increase steadily. By May 2026, AI writers contribute notes to 47.3\% of fact-checked posts. More importantly, 80\% of these posts are not checked by human writers, even though they have been requested by human users. Meanwhile, the duplication rate declines from 38\% in January 2026 to 11\% in March 2026, before rising modestly to 19\% in late April 2026. Together, these results suggest that AI writers continue to evolve over time: they increasingly expand coverage of fact-checking content on \x, mostly to posts unaddressed by human writers, while maintaining relatively low levels of duplication.


\section{Veracity}
\label{sec:veracity}

On Community Notes, submitted notes are not automatically shown to the public. To become visible, a note must be classified as CRH by the platform's bridging algorithm, which requires sufficient ratings and broad agreement among users with different viewpoints. We therefore examine the veracity of AI notes from two complementary perspectives: whether they attract human ratings and whether they are ultimately classified as helpful.

Rating attention is an important precondition for veracity in this setting. Even a high-quality note cannot become publicly visible unless enough users evaluate it, and insufficient attention may delay or prevent the system from determining whether the note is helpful. Thus, if human raters avoid evaluating AI notes, AI fact-checking may be less effective even when the generated notes themselves are accurate or informative. For this reason, we treat rating attention not merely as a measure of engagement, but as part of the evaluation process through which note helpfulness is established.

This evaluation process is especially important for AI-generated notes because they are visibly labeled as such. On the rating interface, AI notes appear with a disclosure stating that they were produced by an experimental AI writer and may contain errors. As a result, human raters evaluate AI notes with explicit knowledge of their source, which may affect both their willingness to rate notes and their assessment of helpfulness. Prior work on algorithm aversion shows that people can lose confidence in algorithmic systems after observing errors, even when those systems perform as well as or better than humans~\cite{dietvorst2015algorithm}. Recent evidence further suggests that AI aversion is stronger when model capability is uncertain and when tasks require contextual or personalized judgment~\cite{qin2025ai}. Community Notes combines both conditions: human raters evaluate context-specific factual claims, while the disclosure label explicitly frames AI notes as experimental and potentially error-prone.

\subsection{Descriptive Veracity Measures}

\begin{figure}[t]
    \centering
    \includegraphics[width=0.6\linewidth]{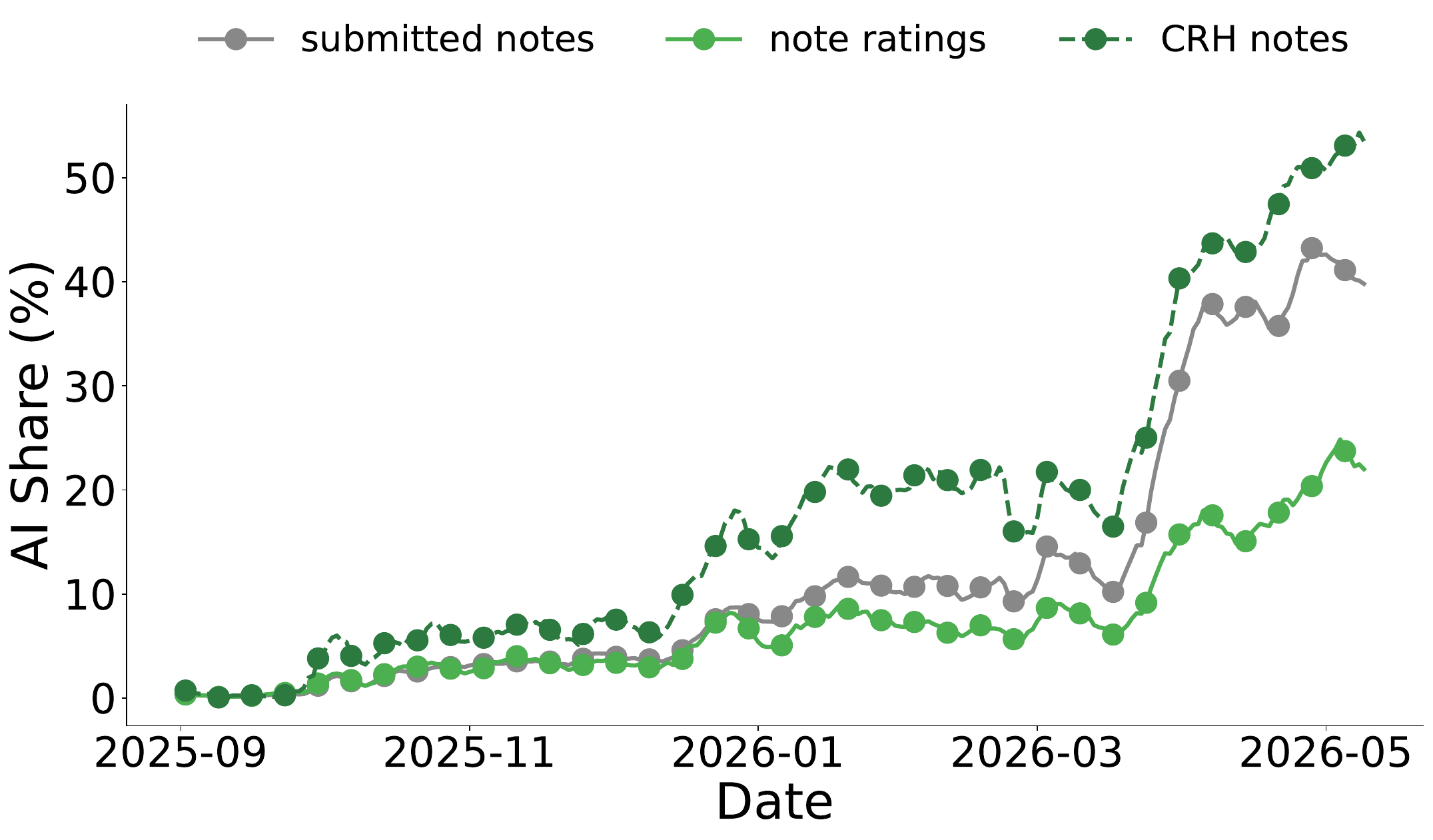}
    \caption{Temporal trends in the AI share of submitted notes, ratings received within 48 hours, and CRH notes. Lines show seven-day rolling averages.}
    \label{fig:veracity_attention_share}
\end{figure}

\Cref{fig:veracity_attention_share} shows temporal trends in the AI share of submitted notes, ratings received within 48 hours, and CRH notes. If AI notes receive proportional attention and achieve proportional helpfulness outcomes, the three lines should closely track one another. Deviations among these lines therefore capture two distinct patterns: the gap between the rating share and the submission share indicates whether AI notes receive more or less attention per submitted note, while the gap between the CRH share and the submission share indicates whether AI notes are classified as helpful at a higher or lower rate than their submission volume would suggest.

Two patterns emerge. First, all three shares increase over time, reflecting the growing participation of AI writers in Community Notes. Second, the AI share of CRH notes is consistently higher than both its share of submitted notes and its share of received ratings. In May 2026, AI writers contribute 41.2\% of submitted notes and 52.2\% of CRH notes, while receiving only 23.1\% of ratings. This suggests that AI notes account for a disproportionately large share of helpful notes despite receiving a disproportionately small share of human ratings. Overall, this result is encouraging: AI notes appear to produce substantial helpful note output while requiring comparatively less human rating efforts.

\begin{figure}[t]
    \centering
    \includegraphics[width=0.6\linewidth]{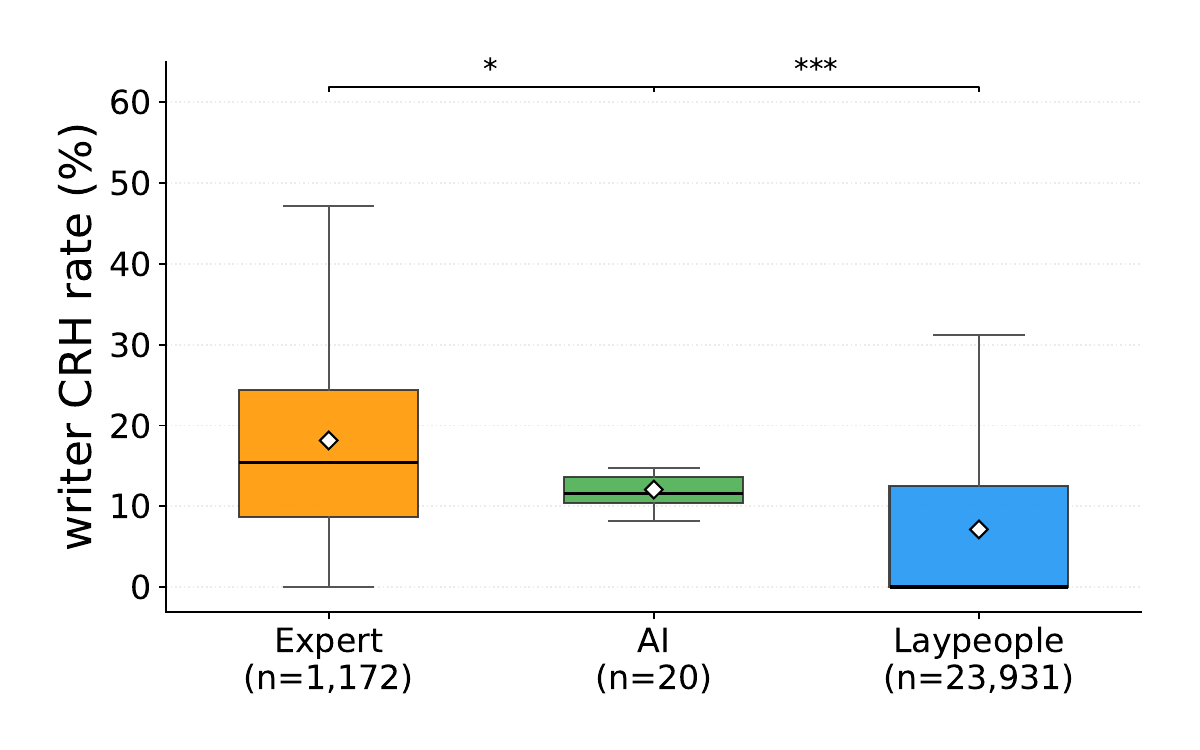}
    \caption{Distributions of writer CRH rates for AI writers, human experts, and laypeople with at least five submitted notes. While circles indicate group means, and $n$ in parentheses indicates the number of writers in each group. Statistical significance is assessed using one-tailed Mann--Whitney U tests. $^{*} p < 0.05, ^{***} p < 0.001$.}
    \label{fig:veracity_writer_crh_rate}
\end{figure}

\Cref{fig:veracity_writer_crh_rate} shows the distribution of writer CRH rates for AI writers, human experts, and laypeople. For each writer, we compute the fraction of submitted notes that eventually reach CRH status. To improve reliability, we restrict this analysis to writers with at least five notes. The mean CRH rate is 12.1\% for AI writers, compared with 18.1\% for human experts and 7.1\% for laypeople. We assess group differences using one-tailed Mann--Whitney U tests, comparing human experts with AI writers and AI writers with laypeople. The results show that human experts have significantly higher CRH rates than AI writers, while AI writers have significantly higher CRH rates than laypeople. This provides initial evidence that AI writers outperform laypeople but still lag behind human experts on Community Notes.

\subsection{Within-Post Veracity Comparisons}

Comparing AI and human notes across different \xs posts may confound differences in attention and helpfulness with post-level factors such as topic, popularity, controversy, and factual complexity. To address this concern, we construct pairs of AI and human notes submitted for the same post, ensuring that both notes address the same fact-checking task.

For each post that receives notes from both AI and human writers, we randomly select one AI note and one human note to form a matched pair. This procedure yields two comparison groups: AI vs. experts ($n$ = 4,981) and AI vs. laypeople ($n$ = 15,190). We estimate 95\% CIs using bootstrapping with 10,000 iterations.

\paragraph{Comparing rating attention}
The first note submitted for a post may receive more ratings simply because it appears earlier. To account for this timing effect, we split each comparison group by submission order, distinguishing pairs where the AI note is submitted earlier from those where the human note is submitted earlier. This yields a $2{\times}2$ design: writer comparison group by submission order. For each condition, we extract the number of ratings received by the AI and human notes in each pair. Because rating counts are unbounded and highly skewed, we measure relative attention as the ratio of ratings received by the AI note to ratings received by the human note. A ratio of one indicates equal rating attention within the pair; values above one indicate more ratings for the AI note, and values below one indicate more ratings for the human note.

\begin{figure}[t]
    \centering
    \includegraphics[width=0.6\linewidth]{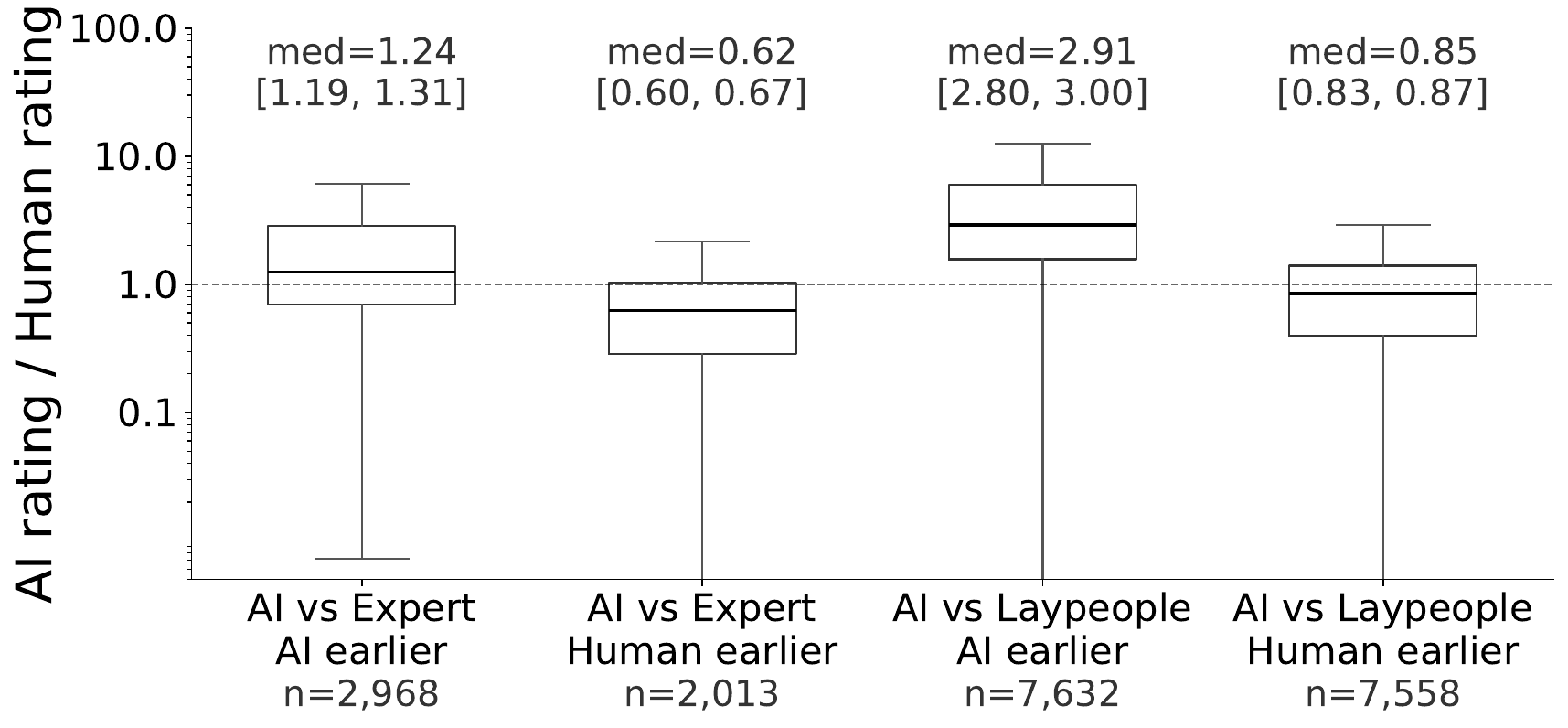}
    \caption{Ratios of ratings received by AI notes relative to matched human notes, controlling for the fact-checked post. Values above one indicate greater rating attention to AI notes, whereas values below one indicate greater rating attention to human notes.
    }
    \label{fig:veracity_pair_attention}
\end{figure}

\Cref{fig:veracity_pair_attention} shows the distribution of rating ratios across the four comparison conditions. When AI notes are submitted earlier, they receive a median of 1.24 (95\% CI: [1.19, 1.31]) times as many ratings as later notes written by experts and 2.91 (95\% CIs [2.80, 3.00]) times as many ratings as later notes written by laypeople. Conversely, when AI notes are submitted later, they receive fewer ratings than earlier human notes: 0.62 (95\% CI: [0.60, 0.67]) times as many ratings as expert-written notes and 0.85 (95\% CI: [0.83, 0.87]) times as many ratings as notes written by laypeople. These patterns suggest a clear first-mover advantage: notes submitted earlier tend to attract more ratings, regardless of whether they are written by AI or humans, although the magnitude of this advantage varies across comparison groups.

\begin{figure}[t]
    \centering
    \includegraphics[width=0.6\linewidth]{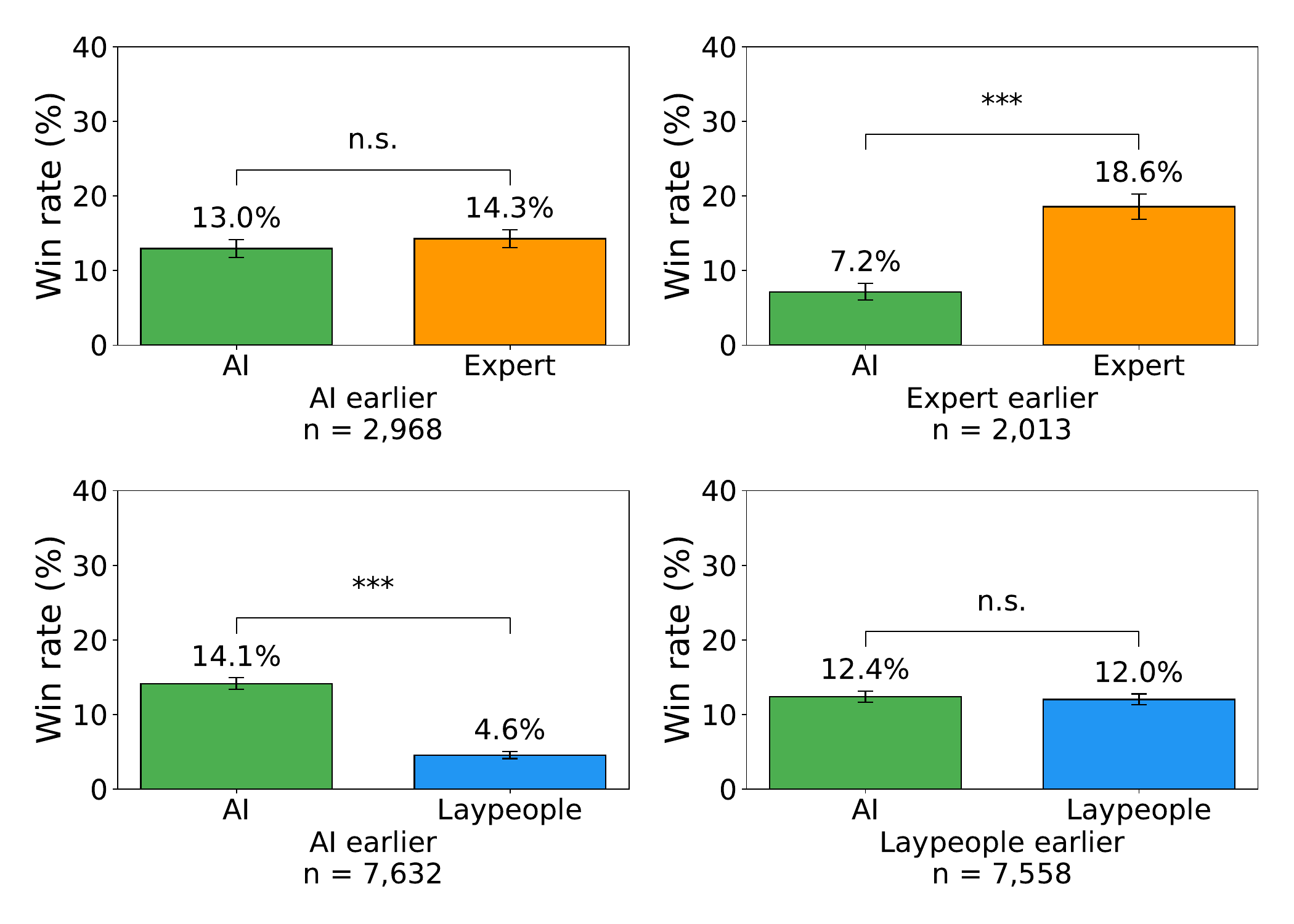}
    \caption{Win rates in head-to-head comparisons between AI and human notes, stratified by writer category and submission order. Top rows: AI vs. experts; bottom rows: AI vs. laypeople. Left columns: AI note submitted earlier; right columns: human note submitted earlier. Statistical significance is assessed using one-tailed Wilcoxon signed-rank tests as the comparisons are paired within posts.}
    \label{fig:veracity_pair_win_rate}
\end{figure}

\paragraph{Comparing note helpfulness}
To isolate note helpfulness from post-level differences and submission-order effects, we conduct a head-to-head comparison within each AI--human note pair. If the AI note is classified as CRH and the human note is not, we count the pair as an AI win. If the human note is CRH and the AI note is not, we count the pair as a human win. If both notes are CRH, or if neither note is CRH, we count the pair as a tie. We then compute AI and human win rates separately for each of the four comparison conditions.

\Cref{fig:veracity_pair_win_rate} shows the results. In comparisons with expert notes, AI notes perform comparably when submitted earlier, with both AI and expert win rates around 13.6\% (top-left panel). However, when expert notes are submitted earlier, they are significantly more likely than later AI notes to reach CRH status, with win rates of 18.6\% for experts and 7.2\% for AI, a difference of 11.4 percentage points (top-right panel). The pattern reverses in comparisons with layperson notes. When AI notes are submitted earlier, they are significantly more likely to reach CRH status than later layperson notes, with win rates of 14.1\% for AI and 4.6\% for laypeople, a difference of 9.5 percentage points (bottom-left panel). However, when layperson notes are earlier, AI and layperson notes perform comparably, with both win rates around 12.2\% (bottom-right panel). Together, these results suggest that AI notes perform comparably to expert notes when AI has the timing advantage but lag behind experts when submitted later. In contrast, AI notes outperform layperson notes when submitted earlier and remain comparable when submitted later.


\section{Operational Heterogeneity and Veracity Homogeneity Among AI Writers}

\begin{figure}[t]
    \centering
    \includegraphics[width=0.6\linewidth]{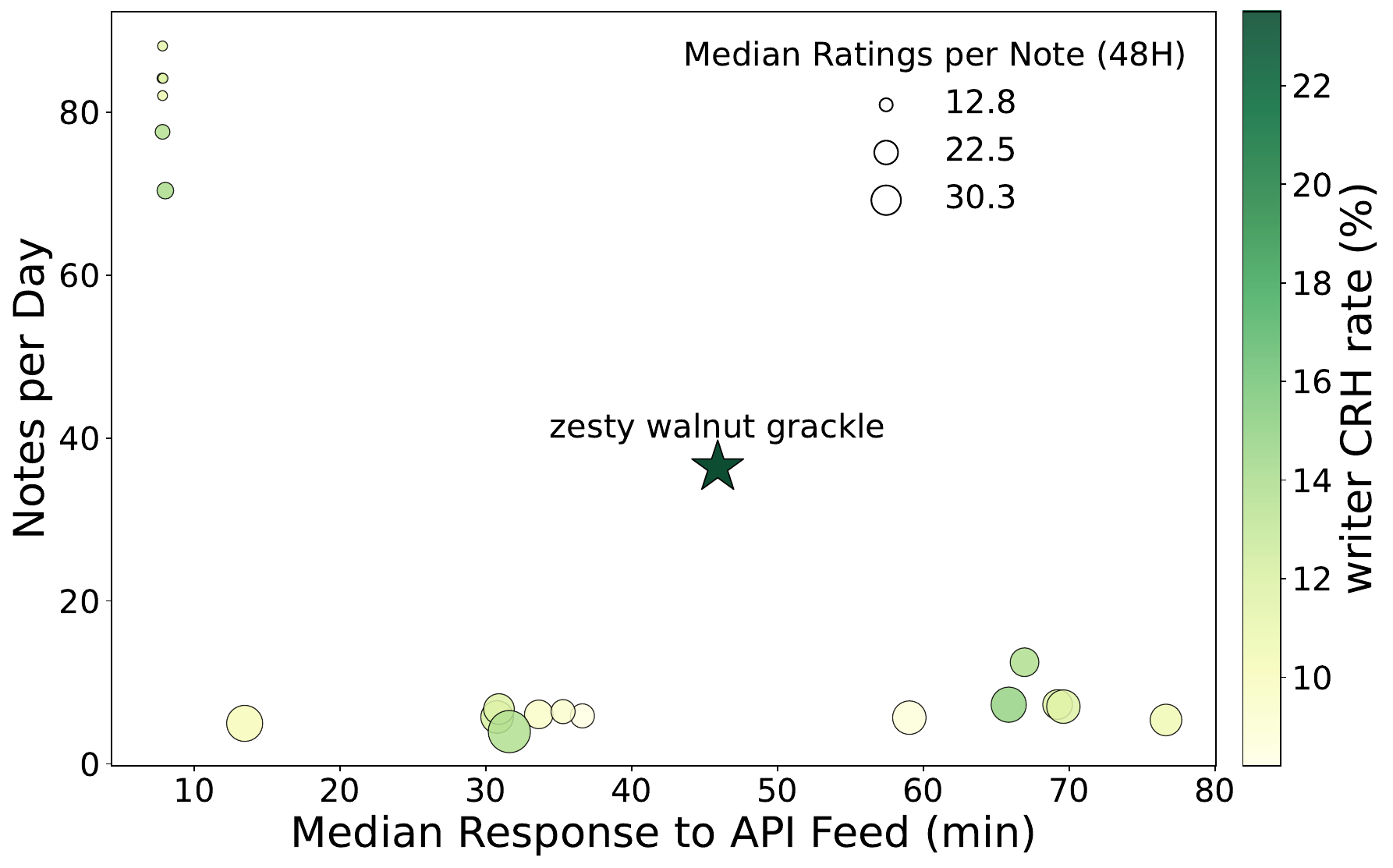}
    \caption{Summary of the 20 AI writers. Each circle represents one writer. The x-axis shows median response time to posts in the API feed, and the y-axis shows mean daily note submissions. Marker size is proportional to median ratings per note within the first 48 hours. Marker color encodes the writer's overall CRH rate, with darker green indicating a higher fraction of notes classified as CRH. The star indicates the only AI writer achieving the highest CRH rate, at 23.5\%.
    }
    \label{fig:case_study}
\end{figure}

To better understand variation across the 20 AI writers, \Cref{fig:case_study} summarizes four dimensions of their attributes: productivity, responsiveness, rating attention, and note helpfulness. The x-axis reports each writer's median response time to posts in the API feed, measured in minutes. The y-axis reports the average number of notes submitted per day during each writer's lifecycle, defined as the period between the writer's first and last active day. Marker size represents the median number of ratings each writer's notes receive within the first 48 hours, while marker color encodes the writer's overall CRH rate. Together, this visualization provides a compact overview of how AI writers differ in activity level, speed, attention, and helpfulness on Community Notes.

We observe both heterogeneity and homogeneity among AI writers. The heterogeneity lies primarily in their operational utility. A small subset of five AI writers, located in the top-left region of~\Cref{fig:case_study}, produces substantially more notes per day and responds much more quickly than the rest. By contrast, most AI writers contribute only sporadically and exhibit more variable response times. This pattern suggests that the impact of AI-driven fact-checking is shaped not merely by the presence of AI writers, but by the operational capacity and deployment intensity of a small number of hyper active systems.

At the same time, AI writers appear to be surprisingly homogeneous in note helpfulness. Only one AI writer (\texttt{zesty walnut grackle}, plotted as a star) reaches a CRH rate of 23.5\%, while the rest have CRH rates concentrated between 8\% and 15\%. This narrow range suggests that, despite large differences in productivity and responsiveness, most AI writers achieve broadly similar helpfulness outcomes. This pattern points to an important distinction between operational heterogeneity and veracity homogeneity: a small number of AI writers are more useful to the Community Notes system because they are deployed more actively, but most do not yet differ markedly in output quality.

One possible explanation is that many AI writers rely on similar design choices, perhaps by following the official implementation guidance provided by \x. If developers adopt comparable prompts, retrieval procedures, and note generation templates, AI writers may converge toward similar outputs. Such homogeneity may limit methodological diversity within the AI Note Writer ecosystem. Rather than reflecting broad experimentation with distinct fact-checking strategies, the current ecosystem may primarily capture variation in usage intensity around a relatively standardized workflow. This suggests that improving AI contributions will require not only broader adoption, but also more substantive innovation in AI fact-checking pipelines for evidence retrieval, claim reasoning, source attribution, and note writing.

\section{Discussion}

Our findings complicate a simple replacement narrative in which automated fact-checkers either outperform or underperform humans as a whole. Instead, AI writers occupy an intermediate position: they are far more productive and responsive than humans and outperform laypeople in note helpfulness, but they still lag behind human experts. This suggests that the relevant comparison is not between AI and humans in general, but between AI and different tiers of human participation. Prior work has compared crowdsourced and professional fact-checking as distinct modes of misinformation governance~\cite{zhao2023insights}; our results add AI as a third mode within this ecosystem. AI appears especially valuable for improving speed, scale, and coverage, while expert judgment remains important for producing the highest-quality contextual notes.

This intermediate position makes human--AI fact-checking both promising and fragile. AI writers can complement humans by rapidly covering posts that might otherwise remain unaddressed. However, AI writers may also compete with humans for attention and visibility. Their speed gives them a first-mover advantage, and their high volume may crowd out human participation if raters or writers increasingly orient around AI-generated notes. Because AI notes do not consistently outperform expert-written notes, platforms should avoid treating AI as a direct substitute for high-performing human contributors. The central challenge is therefore not simply how to combine humans and AI, but how to structure their interaction so that AI expands coverage without displacing the human judgment that gives crowdsourced fact-checking its legitimacy.

This work also raises governance questions about AI-generated fact-checking content. In Community Notes, AI notes are labeled as experimental and potentially error-prone, making their source visible to raters. Prior work on algorithm aversion suggests that people may distrust algorithmic systems after observing errors~\cite{dietvorst2015algorithm}. At the same time, evidence from AI-generated persuasive messages suggests that disclosure labels do not necessarily reduce persuasive effects~\cite{gallegos2025labeling}. We observe a similar tension: AI notes receive less rating attention relative to their submission volume, yet account for a larger share of helpful notes. AI disclosure may therefore shape attention and evaluation without preventing useful AI notes from being judged helpful. Moreover, recent work finds that Community Notes remains vulnerable to rater bias and manipulation~\cite{truong2025community}. High-volume AI participation could introduce new attack surfaces or amplify existing weaknesses. Future systems should preserve transparency while monitoring how AI-generated content affects rating incentives, exposure dynamics, and the resilience of community-based moderation.

\paragraph{Design Implications}
Our findings suggest that future AI writers should be optimized not only for speed and scale, but also for quality. Existing systems already demonstrate clear operational advantages: they submit notes quickly, operate at high volume, and expand coverage to content often left unaddressed by humans. However, their helpfulness remains below that of human experts, and most achieve similar CRH rates. Simply increasing output is therefore unlikely to be sufficient. Future AI fact-checking designs should prioritize stronger evidence retrieval, source attribution, claim verification, and contextual reasoning, so that generated notes are not merely fast, but also well-grounded, non-opinionated, and broadly helpful to raters with diverse viewpoints.

These results also caution against opening the AI Note Writer API broadly without proper governance mechanisms. Expanding access could reduce latency and increase coverage, but if new AI writers rely on similar pipelines or produce outputs of comparable quality, broader participation may mainly increase volume rather than usefulness. It could also intensify competition for rating attention, amplify first-mover advantages, and generate redundant notes on the same posts. The key question is therefore not simply whether more AI writers should be allowed, but how participation should be structured to improve system-level outcomes.

A more sustainable approach is to treat AI writer access as a governed ecosystem rather than an open submission channel. Platforms could require minimum performance thresholds, monitor CRH and duplication rates over time, and allocate visibility or submission privileges based on demonstrated usefulness. They could also reward systems that cover under-addressed posts, use high-quality evidence, or perform well across domains. Under this model, expanding AI participation can be beneficial, but only if paired with safeguards that preserve human participation, limit low-quality or duplicate content, and encourage substantive innovation in AI fact-checking methods.

\paragraph{Ethical considerations}
This study relies exclusively on publicly available Community Notes data and does not involve interaction with human subjects. Generative AI tools were used to assist with writing and programming. The authors retain full responsibility for all analyses, interpretations, and the content of the manuscript.

\paragraph{Limitations and future work}
This study has several limitations. First, our analysis is observational and cannot establish causal effects of AI writers on human participation, rating behavior, or note helpfulness. For example, while AI-generated notes increase rapidly as contributions from laypeople decline, we cannot determine whether AI directly crowds out human contributors or whether both trends reflect broader platform dynamics. Similarly, although earlier notes receive more ratings, observational data cannot fully separate timing effects from unobserved differences in note quality, post salience, or rater behavior. Future work could use field experiments or platform interventions to more directly estimate how AI-generated notes affect human writing and rating dynamics.

Second, our measures of veracity rely on platform-native outcomes, especially ratings and CRH status. These outcomes determine whether notes become visible to general users, but they are not direct measures of factual accuracy: notes may fail to reach CRH because they receive too few ratings, lack cross-partisan consensus, or are posted after stronger competing notes. Conversely, CRH notes may still contain incomplete or imperfect evidence. Future research should combine Community Notes outcomes with independent expert audits, evidence-grounding assessments, or claim-level accuracy evaluations. Because our study focuses on the early deployment period of a small and self-selected AI writer ecosystem, as API rules, disclosure labels, ranking mechanisms, and writer pipelines evolve, future work should also examine how writer diversity, feedback, and platform governance evolve as the system matures.


\section{Conclusion}

This paper presents the first empirical analysis of AI fact-checkers and their contributions to \x's Community Notes. Examining the first eight months of the AI Note Writer API, we characterize AI writers across volume, velocity, variety, and veracity. We find that a small number of AI writers contribute a disproportionately large and rapidly growing share of notes, respond within minutes after posts become available through the API, and expand coverage to many posts not checked by humans. Nonetheless, their veracity remains mixed: both AI and human notes benefit from first-mover advantages for rating attention, and AI-generated notes outperform those written by laypeople but still lag behind those written by experts. These findings suggest that AI writers can substantially improve the speed, scale, and coverage of crowdsourced fact-checking, but should be governed as complements rather than substitutes for human contributors. Our study highlights the need to design human--AI fact-checking systems that preserve human participation, mitigate exposure biases, and ensure that AI contributions are evaluated transparently and fairly.

\bibliographystyle{plainnat}
\bibliography{references}

@inproceedings{allen2022birds,
  title={Birds of a feather don’t fact-check each other: Partisanship and the evaluation of news in Twitter’s Birdwatch crowdsourced fact-checking program},
  author={Allen, Jennifer and Martel, Cameron and Rand, David G},
  booktitle={CHI},
  year={2022}
}

@inproceedings{asikin2026fueling,
  title={Fueling volunteer growth: The case of Wikipedia administrators},
  author={Asikin-Garmager, Eli and Liou, Yu-Ming and Myrick, Caroline and Lo, Claudia and Saez-Trumper, Diego and Zia, Leila},
  booktitle={CHI},
  year={2026}
}

@article{augenstein2024factuality,
  title={Factuality challenges in the era of large language models and opportunities for fact-checking},
  author={Augenstein, Isabelle and Baldwin, Timothy and Cha, Meeyoung and Chakraborty, Tanmoy and Ciampaglia, Giovanni Luca and Corney, David and DiResta, Renee and Ferrara, Emilio and Hale, Scott and Halevy, Alon and others},
  journal={Nature Machine Intelligence},
  year={2024}
}

@article{chuai2026community,
  title={Community-based fact-checking reduces the spread of misleading posts on X (formerly Twitter)},
  author={Chuai, Yuwei and Pilarski, Moritz and Renault, Thomas and Restrepo-Amariles, David and Troussel-Cl{\'e}ment, Aurore and Lenzini, Gabriele and Pr{\"o}llochs, Nicolas},
  journal={Nature Communications},
  year={2026}
}

@inproceedings{de2025supernotes,
  title={Supernotes: Driving consensus in crowd-sourced fact-checking},
  author={De, Soham and Bakker, Michiel A and Baxter, Jay and Saveski, Martin},
  booktitle={TheWebConf},
  year={2025}
}

@article{deverna2024fact,
  title={Fact-checking information from large language models can decrease headline discernment},
  author={DeVerna, Matthew R and Yan, Harry Yaojun and Yang, Kai-Cheng and Menczer, Filippo},
  journal={PNAS},
  year={2024}
}

@article{dietvorst2015algorithm,
  title={Algorithm aversion: People erroneously avoid algorithms after seeing them err},
  author={Dietvorst, Berkeley J and Simmons, Joseph P and Massey, Cade},
  journal={Journal of Experimental Psychology: General},
  year={2015}
}

@article{gallegos2025labeling,
  title={Labeling messages as AI-generated does not reduce their persuasive effects},
  author={Gallegos, Isabel O and Shani, Chen and Shi, Weiyan and Bianchi, Federico and Gainsburg, Izzy and Jurafsky, Dan and Willer, Robb},
  journal={arXiv preprint arXiv:2504.09865},
  year={2025}
}

@article{gong2026effects,
  title={The effects of request alerts on the diversity and visibility of community notes},
  author={Gong, Yilin and Wu, Siqi},
  journal={arXiv preprint arXiv:2604.17042},
  year={2026}
}

@article{li2025scaling,
  title={Scaling human judgment in Community Notes with LLMs},
  author={Li, Haiwen and De, Soham and Revel, Manon and Haupt, Andreas and Miller, Brad and Coleman, Keith and Baxter, Jay and Saveski, Martin and Bakker, Michiel},
  journal={Journal of Online Trust and Safety},
  year={2025}
}

@article{li2026ai,
  title={AI fact-checking in the wild: A field evaluation of LLM-written community notes on X},
  author={Li, Haiwen and Bakker, Michiel A},
  journal={arXiv preprint arXiv:2604.02592},
  year={2026}
}

@inproceedings{lloyd2026beyond,
  title={Beyond community notes: A framework for understanding and building crowdsourced context systems for social media},
  author={Lloyd, Travis and Nguyen, Tung and Levy, Karen and Naaman, Mor},
  booktitle={CHI},
  year={2026}
}

@article{mohammadi2025ai,
  title={AI feedback enhances community-based content moderation through engagement with counterarguments},
  author={Mohammadi, Saeedeh and Yasseri, Taha},
  journal={arXiv preprint arXiv:2507.08110},
  year={2025}
}

@article{moon2025fact,
  title={Fact-checking in the age of AI: Reducing biases with non-human information sources},
  author={Moon, Won-Ki and Kahlor, Lee Ann},
  journal={Technology in Society},
  year={2025}
}

@article{qin2025ai,
  title={AI aversion or appreciation? A capability--personalization framework and a meta-analytic review},
  author={Qin, Xin and Zhou, Xiang and Chen, Chen and Wu, Dongyuan and Zhou, Hansen and Dong, Xiaowei and Cao, Limei and Lu, Jackson G},
  journal={Psychological Bulletin},
  year={2025}
}

@inproceedings{roitero2025efficiency,
  title={Efficiency and effectiveness of LLM-based summarization of evidence in crowdsourced fact-checking},
  author={Roitero, Kevin and Wright, Dustin and Soprano, Michael and Augenstein, Isabelle and Mizzaro, Stefano},
  booktitle={SIGIR},
  year={2025}
}

@article{sahnan2026can,
  title={Can LLMs automate fact-checking article writing?},
  author={Sahnan, Dhruv and Corney, David and Larraz, Irene and Zagni, Giovanni and Miguez, Ruben and Xie, Zhuohan and Gurevych, Iryna and Churchill, Elizabeth and Chakraborty, Tanmoy and Nakov, Preslav},
  journal={TACL},
  year={2026}
}

@article{singh2026gitsearch,
  title={GitSearch: Enhancing community notes generation with gap-informed targeted search},
  author={Singh, Sahajpreet and Jaidka, Kokil and Kan, Min-Yen},
  journal={arXiv preprint arXiv:2602.08945},
  year={2026}
}

@article{truong2025delayed,
  title={Delayed takedown of illegal content on social media makes moderation ineffective},
  author={Truong, Bao Tran and Kim, Sangyeon and Nogara, Gianluca and Verdolotti, Enrico and Sahneh, Erfan Samieyan and Saurwein, Florian and Just, Natascha and Luceri, Luca and Giordano, Silvia and Menczer, Filippo},
  journal={arXiv preprint arXiv:2502.08841},
  year={2025}
}

@article{truong2025community,
  title={Community Notes are vulnerable to rater bias and manipulation},
  author={Truong, Bao Tran and Wu, Siqi and Flammini, Alessandro and Menczer, Filippo and Stewart, Alexander J},
  journal={arXiv preprint arXiv:2511.02615},
  year={2025}
}

@article{wojcik2022birdwatch,
  title={Birdwatch: Crowd wisdom and bridging algorithms can inform understanding and reduce the spread of misinformation},
  author={Wojcik, Stefan and Hilgard, Sophie and Judd, Nick and Mocanu, Delia and Ragain, Stephen and Hunzaker, MB and Coleman, Keith and Baxter, Jay},
  journal={arXiv preprint arXiv:2210.15723},
  year={2022}
}

@inproceedings{wolfe2024impact,
  title={The impact and opportunities of generative AI in fact-checking},
  author={Wolfe, Robert and Mitra, Tanushree},
  booktitle={FAccT},
  year={2024}
}

@inproceedings{wu2020variation,
  title={Variation across scales: Measurement fidelity under twitter data sampling},
  author={Wu, Siqi and Rizoiu, Marian-Andrei and Xie, Lexing},
  booktitle={ICWSM},
  year={2020}
}

@inproceedings{wu2026beyond,
  title={Beyond the crowd: LLM-augmented community notes for governing health misinformation},
  author={Wu, Jiaying and Fu, Zihang and Wang, Haonan and Li, Fanxiao and Guo, Jiafeng and Nakov, Preslav and Kan, Min-Yen},
  booktitle={ACL},
  year={2026}
}

@inproceedings{xing2025evaluating,
  title={Evaluating evidence attribution in generated fact checking explanations},
  author={Xing, Rui and Baldwin, Timothy and Lau, Jey Han},
  booktitle={NAACL},
  year={2025}
}

@misc{young2025ai,
  title = {World's first AI Community Note},
  author = {Young, Nathan},
  year = {2025},
  howpublished = {\url{https://nathanpmyoung.substack.com/p/worlds-first-ai-community-note}}
}

@inproceedings{yang2026fact,
  title={A fact-checking framework with denoising evidence retrieval and LLM-based debate verification},
  author={Yang, Jun and Bai, Yuhan and Song, Dandan and Wu, Zhijing and Tian, Yuhang},
  booktitle={TheWebConf},
  year={2026}
}

@article{zhang2025commenotes,
  title={Commenotes: Synthesizing organic comments to support community-based fact-checking},
  author={Zhang, Shuning and Wang, Linzhi and Shi, Dai and Chuai, Yuwei and Chen, Jingruo and Chen, Yunyi and Wang, Yifan and Wang, Yating and Yi, Xin and Li, Hewu},
  journal={arXiv preprint arXiv:2509.11052},
  year={2025}
}

@article{zhao2023insights,
  title={Insights from a comparative study on the variety, velocity, veracity, and viability of crowdsourced and professional fact-checking services},
  author={Zhao, Andy and Naaman, Mor},
  journal={Journal of Online Trust and Safety},
  year={2023}
}

@article{zhou2024correcting,
  title={Correcting misinformation on social media with a large language model},
  author={Zhou, Xinyi and Sharma, Ashish and Zhang, Amy X and Althoff, Tim},
  journal={arXiv preprint arXiv:2403.11169},
  year={2024}
}

\end{document}